# Specialization vs diversification in research activities: the extent, intensity and relatedness of field diversification by individual scientists[1]

*Giovanni Abramo*[a,*], *Ciriaco Andrea D'Angelo*[a,b], *Flavia Di Costa*[a]

[a] Laboratory for Studies in Research Evaluation
Institute for System Analysis and Computer Science (IASI-CNR)
National Research Council of Italy

[b] Department of Engineering and Management
University of Rome "Tor Vergata"

## Abstract

We investigate whether and in what measure scientists tend to diversify their research activity, and if this tendency varies according to their belonging to different disciplinary areas. We analyze the nature of research diversification along three dimensions: extent of diversification, intensity of diversification, and degree of relatedness of topics in which researchers diversifies. For this purpose we propose three bibliometric indicators, based on the disciplinary placement of scientific output of individual scientists. The empirical investigation shows that the extent of diversification is lowest for scientists in Mathematics and highest in Chemistry; intensity of diversification is lowest in Earth sciences and highest in Industrial and information engineering; and degree of relatedness is lowest in Earth sciences and highest in Chemistry.

[*] *Corresponding author*

# 1. Introduction

The scientific "disciplines" identify specific means of structuring the reality of knowledge, while knowledge itself maintains its character as a whole, since all the sciences are nothing other than the product of a single human intellectual activity. Albert Einstein asserted that science itself is a "creation" of the human intellect. The disciplinary structure of science is a very recent artifact. According to Klein (1996), the modern concept of the scientific discipline is only a few centuries old, experiencing slow evolution through the 1700s, but then becoming more rapid. This was due above all to the development of specialization of labor in the academic environment starting in the 19$^{th}$ century. This specialization, although it has contributed to the boundless development of knowledge, has also created ever more pronounced disciplinary boundaries, and so continual isolation of the fields of knowledge themselves and of the communities dedicated to them (Boulding, 1956).

The birth of the discipline of bibliometrics, in the 1960s, reflected the exponential growth of all scientific production. However it is no accident that on the other hand, it also saw the introduction of schemes and standards of disciplinary classification, now universally applied (Price, 1963).

"Interdisciplinary research" (IDR) is in fact predicated on disciplinarity, meaning mastery of the individual disciplines, including knowledge of their logical and methodological structure. Such knowledge is necessary to identify possible interactions between fields that could unite in understanding a phenomenon or find solutions for specific problems. Schmidt (2008) views IDR as "a highly valued tool in order to restore the unity of sciences or to solve societal-pressing problems". Rhoten, Caruso & Parker (2003) consider that it is the nature of problems being addressed, particularly their complexity, that leads to the rise in IDR, and of scientists' attempts to cross disciplinary boundaries. These are often erected by the scientific communities themselves to respond to needs of governance.

The development of the IDR phenomenon immediately attracted the attention of many scholars and raised challenges on various fronts. The issues receiving most attention are the taxonomic problem, the development of schemes for conceptual and practical definition of IDR, and subsequently its measurement (Huutoniemi, Klein, Bruun, & Hukkinen, 2010). According to a review by Klein (2008), the most commonly accepted scheme for definition of IDR is that involving three concepts: multidisciplinarity, interdisciplinarity and transdisciplinarity. Each of these is characterized by a particular type of "knowledge integration", meaning the particular type of merging of theories and concepts, techniques and tools, information and data, from various fields of knowledge (Porter, Roessner, Cohen, & Perreault, 2006). According to Wagner et al. (2011) the phenomenon of knowledge integration can occur within a single mind, as well as within teams.

The above introduces the general current of studies on IDR. Within this, the aim of the present work is to analyze the dichotomy between specialization and diversification in scientists' research activity. What we want to investigate is whether and in what measure scientists tend to diversify their research activity, and if this tendency varies according to their belonging to different disciplinary areas. For this purpose we analyze the disciplinary placement of scientific output of individual scientists, based on bibliometric techniques. Although such techniques cannot inform on the social aspects at the basis of knowledge integration, they do offer valid support (quantitative, objective, and also quite economical) for the radiography and better understanding the phenomenon. Empirical investigation requires the construction of the publication portfolio of each individual scientist, from which we can proceed to disciplinary classification of the works. In the case of Italian academics, we can apply an algorithm for disambiguation of author names, developed by D'Angelo, Giuffrida & Abramo (2011), which allows us to assign the publications indexed in the Web of Science (WoS) to their relative academic authors. Using the WoS classification scheme, each of these publications is also assigned to one or more subject categories, depending on the classification of the hosting journal. In this way we can analyze the nature of research diversification for each academic. Furthermore, exploiting a unique characteristic of the Italian academic system, whereby each academic is classified in one and only one field of research, we can assess whether the nature of diversification varies across fields.

The presentation of the work is as follows. In Section 2 we provide a summary of literature on the theme of measuring interdisciplinarity; in the next we illustrate the specific methods of our empirical investigation, in terms of indicators and dataset. Section 4 presents the analytical results in three parts, for each of the three dimensions/indicators considered. Section 5 concludes the work with the author's considerations.

## 2. Literature review

According to Stirling (1994), IDR displays some combination of only three basic properties, named "variety, "balance" and "disparity". Stirling (2007) also proposes the indicators suited to measurement of each: the "variety" indicator is defined as the answer to the question: "how many types of thing do we have?"; for "balance" it is instead the answer to "how much of each type of thing do we have?"; finally, "disparity" is "how different from each other are the types of thing that we have?".

In the bibliometric sphere, these concepts have been widely applied in the investigation of IDR, as demonstrated in a review of the issue by Wagner et al. (2011). Many studies take a bottom-up approach, building from measurement of interdisciplinarity for individual articles. The proposed measures are based on the disciplinary profile of the references cited, considering that reference to the preceding literature in various disciplines is as a signal of acquisition and integration of the results of these disciplines (Porter, Cohen, Roessner & Perreault, 2007; Rafols & Meyer, 2010; Wang, Thijs & Glänzel, 2015; Mugabushaka, Kyriakou & Papazoglou, 2016). In particular, Porter & Rafols (2009) used the works published in a cluster of selected journals indexed in the WoS over the period 2007-2011, examining their relative lists of references and identifying the disciplinary areas of the works cited, in terms of: i) number of subject categories (SCs) cited; ii) distribution of the citations among the SCs; iii) similarity or disparity among these SCs. In substance, Porter & Rafols proposed the measurement of Stirling's (1994) three basic properties of research diversification through mapping the subject categories of cited publications. Zhang, Rousseau & Glänzel (2016) adopt the same approach to study the interdisciplinarity of journals.

Other studies are instead based on a top-down approach: again using typical disciplinary classifications such as WoS subject categories, they study the frequency distributions for scientific portfolios produced by defined units of analysis. For example, van Raan & van Leeuwen (2002) propose an approach for measurement of the IDR of a research organization through the percentage of its publications in each SC, or of citations received from each SC. Bourke & Butler (1998) had previously investigated the IDR conducted within Australian university departments, through the analysis of journals hosting the 1990-1994 publications authored by researchers of each department. Rinia et al. (2001) applied a similar approach in the Netherlands, analyzing the outcomes from a nation-wide evaluation program of all academic groups in physics. As an aside, the intention of the last two works was to understand if IDR should be assessed in the same way as "disciplinary" research.

One of the areas of analysis little visited by scholars concerns IDR at the level of individual researcher. This is a strategic area, if we think of the challenges of complex research and the current recourse to policy aimed at incentivizing interdisciplinary work and thus influencing choices by the protagonists – the researchers themselves. The only contribution in the literature seems to be that from Schummer (2004), who carried out a coauthor analysis of nanotech journals in 2002-2003. By mapping "disciplinary" affiliation of coauthors, he was able to measure the IDR of each scientist in terms of interaction with the disciplines associated to all their coauthors. Using a similar approach, Abramo et al., (2012) analyzed the degree of collaboration among scientists from different disciplines in order to identify the most frequent "combinations of knowledge" in research activity, drawing on 2004-2008 WoS publications by all Italian professors in the sciences.

Following on this preceding work, the authors now intend to study the scientists' propensity to diversify, through the disciplinary placement of their scientific production.

## 3. Methodology

### 3.1 Dataset

The dataset for the analysis is the 2004-2008[2] scientific production achieved by Italian professors in the sciences. In the Italian academic system all professors are classified in one and only one field (named "scientific disciplinary sector", or SDS, of which 370 in all), grouped into disciplines (named "university disciplinary areas", UDAs, 14 in all).[3] In this study we focus on the sciences, for which the WoS coverage of publications by Italian universities is satisfactory. The sciences consist of 192 SDSs grouped into nine UDAs.[4]

Data on academics are extracted from a database maintained at the central level by the Ministry of Education, University and Research[5], indexing the name, academic rank, affiliation, and the SDS of each professor. Publication data are drawn from the Italian Observatory of Public Research (ORP), a database developed and maintained by the authors and derived under license from the WoS. Beginning from the raw data of Italian publications indexed in the WoS, we apply a complex algorithm for disambiguation of the true identity of the authors and their institutional affiliations (for details see D'Angelo, Giuffrida, & Abramo, 2011).

---

[2] The choice of a publication window quite far in the past is in consideration of a planned follow-up study with the aim of assessing whether interdisciplinary output is more influential in terms of citations: a longer citation window assures more robust and reliable results.

[3] For the complete list see http://attiministeriali.miur.it/UserFiles/115.htm, last accessed on December 23, 2016.

[4] Mathematics and computer sciences; Physics; Chemistry; Earth sciences; Biology; Medicine; Agricultural and veterinary sciences; Civil engineering; Industrial and information engineering.

[5] See http://cercauniversita.cineca.it/php5/docenti/cerca.php, last accessed on December 23, 2016.

The overall dataset is composed of 33,784 publishing professors. Table 1 shows their division by UDA, as well as the relative scientific production[6] for the five-year period under observation.

*Table 1: Dataset of the analysis*

| UDA | SDSs | Professors | Publications |
|---|---|---|---|
| 1 - Mathematics and computer science | 10 | 2,814 | 15,049 |
| 2 - Physics | 8 | 2,605 | 24,505 |
| 3 - Chemistry | 12 | 3,391 | 24,923 |
| 4 - Earth sciences | 12 | 1,161 | 4,996 |
| 5 - Biology | 19 | 5,160 | 28,881 |
| 6 - Medicine | 50 | 10,100 | 57,971 |
| 7 - Agricultural and veterinary sciences | 30 | 2,745 | 10,765 |
| 8 - Civil engineering | 9 | 1,151 | 4,543 |
| 9 - Industrial and information engineering | 42 | 4,657 | 34,496 |
| Total | 192 | 33,784 | 182,675* |

*\* The total is less than the sum of column data due to multiple counting of individual publications authored by professors of more than one UDA.*

### 3.2 Indicators

The disciplines in which research activity is classified often overlap, and generally have quite weak boundaries – the fields within them even more so. The confines are anything but static, since continuous scientific progress contributes to variation in the scope of the fields, as well as the birth of new ones and disappearance of old ones. This said, the analysis of IDR must in any case involve some predetermined reference classification. For our study we use the WoS classification system. We associate each publication in the database with only one topic. By topic we mean the SC of the hosting journal in the case of a mono-category journal, or the combination of WoS SCs when the publication is issued in a multi-category journal.

The authors can thus be divided into two classes: those who diversify, meaning their publications fall in more than one topic; those who do not diversify, meaning their publications fall in a single topic. We refer to these classes as "diversified" and "specialized" authors. Obviously the distribution of the scholars between the two classes depends on the breadth of the publication window observed, as well as the classification scheme for disciplines. The object of our study is the diversified authors. For each, we can first of all identify the dominant topic in which the individual works, meaning the most recurrent SC or SC combination in their publication portfolio. We consider the case of Mario Rossi (John Doe in English), professor in FIS/03 (Physics of matter), who in the period of observation produced eight articles published in four different journals (*Physical Review B*, *Physical Review E*, *Chemphyschem* and *Physical Review letters*). Given the classification of these journals under the WoS system, we have the distribution illustrated in Table 2. The eight articles fall in four different topics, of which the dominant one is subject category UK (Physics, condensed matter), given that half of Rossi's publications fall in this topic. We can also observe cases of more than one dominant topic, which above all is more likely when the subject's number of publications is low.

---
[6] Article, reviews, letters and conference proceedings

*Table 2: Publication portfolio of a professor in the dataset*

| Topic | Discipline | N. of publications | WoS_ID |
|---|---|---|---|
| UK (Physics, condensed matter) | Physics | 4 | 243195800122; 245330200070; 260574500061;251986500011 |
| UF+UR (Physics, fluids & plasmas; Physics, mathematical) | Physics | 2 | 228818200106; 242408800041 |
| EI+UH (Chemistry, physical; Physics, atomic, molecular & chemical) | Chemistry; Physics | 1 | 231971100043 |
| UI (Physics, multidisciplinary) | Physics | 1 | 229700800052 |

We will investigate three dimensions of diversification of research by individual professors. The first is "*extent of diversification (ED)*", measured by an indicator of the same name, given by the number of topics covered in the person's scientific portfolio. The second is intensity of diversification, meaning what share of the researcher's output falls outside of their sector of specialization – measured by the indicator "*diversification ratio (DR)*", given by the ratio of the share of papers falling in topics other than the dominant one to the total number of publications. The higher the value of ED and the closer DR is to one, the more the individual's research activity is diversified. The opposite situation denotes a highly specialized researcher. There can also be antithetical situations: i) a high ED value jointly with a low DR value would indicate that the subject is predominantly specialized but open to exploring new fields; ii) a low ED value with a DR value tending to one is quite unlikely, unless the scientific production is very low. The last dimension investigated is the cognitive relatedness of the topics studied by the academic. Measurement of this requires definition of a threshold of proximity. For this purpose we associate the individual WoS topics to the disciplines,[7] meaning we can identify the topics as "related" if they fall within the same discipline. The indicator for this dimension is "*relatedness ratio (RR)*", equal to the ratio of number of papers in the dominant discipline to total number of papers. An RR of 1 indicates that the researcher, although diversifying, does not go beyond their own disciplinary area. It is likely that a statistician (whose sphere of research can range from statistics to economy, medicine, agriculture, etc.) would have a much lower degree of relatedness than a surgeon.

According to the above taxonomy, for Mario Rossi we observe:
- An *ED* of 4;
- A *DR* of 0.5, since half of the total publications fall outside his dominant topic (Physics, condensed matter);
- An *RR* of 7/8, since 7 of the 8 publications are associated with the dominant discipline (Physics).

---

[7] Each WoS subject category is associated with a single discipline, i.e., one of: Mathematics; Physics; Chemistry; Earth and space sciences; Biology; Biomedical research; Clinical medicine; Psychology; Engineering; Economics; Law, political and social sciences; Multidisciplinary sciences; Art and humanities.

## 4. Analysis

In this section we analyze the three main dimensions characterizing research diversification, by the above indicators ED, DR and RR. The analysis is conducted for each individual academic and then brought to the SDS and UDA levels. For reasons of space we present only exemplary cases along each dimension. All others are reported in the Supplementary Material (SM).

### 4.1 Extent of research diversification

By definition, a researcher's ED can vary between 1 and n-1, where n is the number of topics in which WoS classifies the journals of publication. The maximum value will in reality be much lower, since a single individual could not successfully function in the entire spectrum of knowledge and competencies covered by WoS journals.

As an example of analysis, Table 3 reports some descriptive statistics of ED for professors belonging to the SDSs of Physics UDA. The percentage of diversified professors varies from a minimum of 52% in FIS/08 (Didactics and history of physics) to maximum of 96% in FIS/03 (Physics of matter) and FIS/04 (Nuclear and subnuclear physics). Five SDSs of eight show more than 80% of professors who diversify. The mean ED of the professors in each SDS is shown in column 4 of Table 3. In the top three positions, with quite similar values, are: FIS/07 (Applied physics, 7.2), FIS/03 (Physics of matter, 7.1) and FIS/01 (Experimental physics, 6.6). At the bottom we find FIS/08 (Didactics and history of physics, 2.5). This is also the SDS with the lowest value of min-max range for number of topics. In all SDSs, the extreme low of range is always 1, while the extreme high arrives at a remarkable 58, in FIS/01 (Experimental physics), followed by FIS/03 (Physics of matter, 34) and FIS/07 (Applied physics, 33).

Since the number of topics in which a professor publishes, and thus their ED, is a function reasonably increasing with number of publications produced, in the last column of Table 3 we report the mean value of ratio of ED to number of publications for the given SDS. The values vary in the interval 0.2÷0.5, with averages greater than 0.4 for, FIS/07 (Applied physics, 0.49), FIS/06 (Physics for earth and atmospheric sciences, 0.47), and FIS/08 (Didactics and history of physics, 0.41).

Figure 1 provides the scatter plot of correlation between ED and number of publications for the 435 diversified professors of FIS/03 (Physics of matter). As we would expect, the Pearson $\rho$ correlation coefficient between the two measures is high and positive (0.78); still, dividing the plot in quadrants delimited by the average values of the two measures considered (7.1 vs 20.9), we can observe subpopulations with distinct characteristics. Obviously there is a strong concentration in the lower left quadrant, populated by 218 (50%) professors, characterized by below-average values for both measures. At the same time, the 132 (30%) professors in the upper right quadrant, having a high number of publications, also show dispersion of these among a high number of topics – typical of activity in highly interdisciplinary research. We also note the presence of 51 (12%) professors in the lower right quadrant, who publish a great deal yet still diversify little. The opposite case is represented by 34 (8%) professors in the upper left quadrant – although these publish little, they are notably differentiated (ED equal to or greater than 8).

*Table 3: Extent of research diversification (ED) for professors in the SDSs of Physics: descriptive statistics*

| SDS† | No of professors | % of diversified professors | Average ED | Min-Max ED | Average No of publications | Average ED per publication |
|---|---|---|---|---|---|---|
| FIS/01 | 1003 | 94% | 6.6 | 1-58 | 37.9 | 0.34 |
| FIS/02 | 350 | 89% | 4.4 | 1-18 | 14.9 | 0.35 |
| FIS/03 | 455 | 96% | 7.1 | 1-34 | 20.9 | 0.39 |
| FIS/04 | 165 | 96% | 5.0 | 1-19 | 25.0 | 0.30 |
| FIS/05 | 186 | 70% | 3.1 | 1-15 | 23.4 | 0.21 |
| FIS/06 | 71 | 75% | 4.5 | 1-14 | 12.2 | 0.47 |
| FIS/07 | 344 | 91% | 7.2 | 1-33 | 17.9 | 0.49 |
| FIS/08 | 31 | 52% | 2.5 | 1-5 | 9.0 | 0.41 |

† *FIS/01, Experimental physics; FIS/02, Theoretical physics, Mathematical models and methods; FIS/03, Material physics; FIS/04, Nuclear and subnuclear physics; FIS/05, Astronomy and astrophysics; FIS/06, Physics for earth and atmospheric sciences; FIS/07, Applied physics (cultural heritage, environment, biology and medicine); FIS/08, Didactics and history of physics*

*Figure 1: Relation between extent of diversification and number of publications of professors (435 in all) in FIS/03 (Physics of matter)*

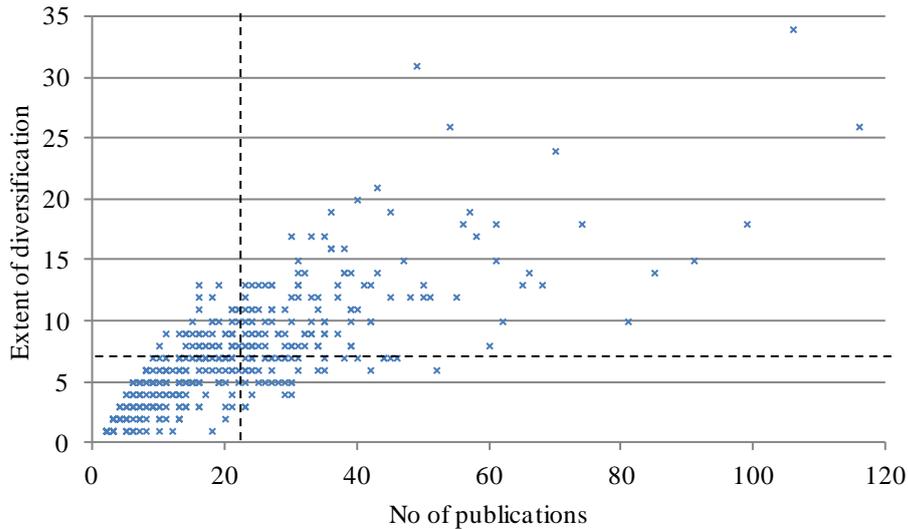

The same ED/number of publications relation is analyzed for FIS/01 (Experimental physics) (Figure 2). Here we observe a very low correlation (Pearson ρ coefficient = 0.17), due to a cluster of some tens of professors who publish a great deal but diversify little. The case of Roberto Cingolani, director of the Italian Institute of Technology is unusual: this is an individual who both publishes and diversifies a great deal, evidently an effect of collaborating with the organization's numerous research groups.

Table 4 presents the descriptive statistics for the SDSs of each UDA.[8] The percentage of diversified professors is highest in Chemistry (average 92%); lowest in Civil engineering (71%). The maximum variation in range (max-min) is observed for Agricultural and veterinary sciences (from 33% in AGR/01-Rural economy and evaluation to 95% in VET/02-Veterinary physiology), while the minimum is seen in Chemistry (90% in CHIM/12-Environmental chemistry and chemistry for cultural heritage, to 97% in CHIM/11-Chemistry and biotechnology of fermentations). This indicates a great variability of diversification behavior among SDSs of Agricultural and veterinary sciences relative to greater uniformity among those of Chemistry. Continuing the examination of minimum values, there are only two UDAs showing an SDS with percentage of professors who diversify at less than 50% − these are AGR/01 for UDA 7, and ICAR/04 (Road, railway and airport construction) for UDA 8. Instead, the maximums for share of diversified professors are consistently above 80%.

---

[8] For reasons of significance, we omit the SDSs (8 in all) with less than 10 diversified professors.

Concerning *ED*, we observe averages per SDS ranging from 3.1 in Civil engineering to 6.0 in Chemistry. The maximum range (min-max) within an individual UDA is observed in Industrial and information engineering (from 1.7 in ING-IND/01-Naval architecture to 8.8 in ING-IND/34-Industrial bioengineering). The greatest uniformity is found in Earth sciences, a UDA where the average ED of professors varies between the minimum of 2.4 in GEO/02-Stratigraphic and sedimentological geology, and maximum 4.3 in GEO/06-Mineralogy. The complete data of ED for all the SDSs and UDAs are provided in the supplementary material (SM1).

*Figure 2: Relation between the extent of diversification and number of publications of professors (947 in all) in FIS/01 (Experimental physics)*

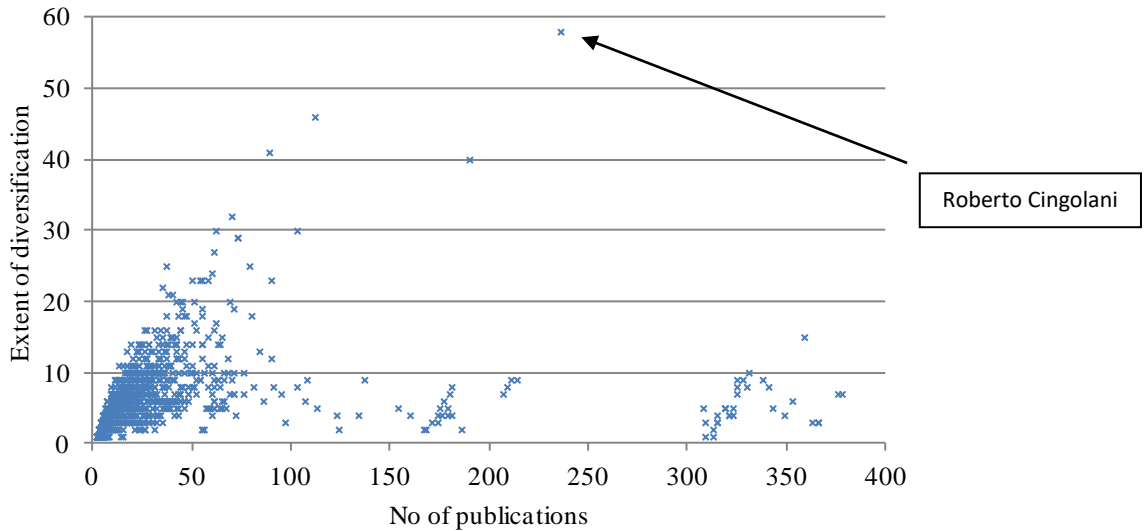

*Table 4: Descriptive statistics of extent of diversification in SDSs of each UDA*

| | % of diversified professors | | Extent of diversification | |
|---|---|---|---|---|
| UDA† | Average | Min-Max | Average | Min-Max |
| 1 | 75 | 51 (MAT/04)-91 (MAT/09) | 3.5 | 1.4 (MAT/02)-5.9 (INF/01) |
| 2 | 83 | 52 (FIS/08)-96 (FIS/04) | 5.0 | 2.5 (FIS/08)-7.2 (FIS/07) |
| 3 | 92 | 90 (CHIM/12)-97 (CHIM/11) | 6.0 | 5.0 (CHIM/09)-7.3 (CHIM/07) |
| 4 | 79 | 55 (GEO/05)-88 (GEO/06) | 3.3 | 2.4 (GEO/02)-4.3 (GEO/06) |
| 5 | 87 | 63 (BIO/02)-94 (BIO/15) | 5.0 | 2.6 (BIO/02)-7.2 (BIO/15) |
| 6 | 81 | 62 (MED/43)-95 (MED/15) | 5.0 | 2.6 (MED/30)-8.6 (MED/01) |
| 7 | 77 | 33 (AGR/01)-95 (VET/02) | 3.6 | 1.7 (VET/09)-5.7 (VET/06) |
| 8 | 71 | 46 (ICAR/04)-82 (ICAR/08) | 3.1 | 2.0 (ICAR/05)-4.7 (ICAR/08) |
| 9 | 84 | 57 (ING-IND/28)-98 (ING-IND/34) | 5.0 | 1.7 (ING-IND/01)-8.8 (ING-IND/34) |
| Total | 81 | 33 (AGR/01)-98 (ING-IND/34) | 4.6 | 1.4 (MAT/02)-8.8 (ING-IND/34) |

*† 1, Mathematics and computer science; 2, Physics; 3, Chemistry; 4, Earth sciences; 5, Biology; 6, Medicine; 7, Agricultural and veterinary sciences; 8, Civil engineering; 9, Industrial and information*

*engineering*

**4.2 Intensity of research diversification**

The analysis of extent of diversification, as above, can be usefully supplemented by assessing intensity of diversification. Therefore, in this section we analyze the intensity of diversification of professors at the SDS and UDA levels, using the indicator *DR*. Returning to our example of the Physics UDA, we calculate the *DR* of professor Cingolani, outlier in Figure 2, author of 236 publications spreading over 59 topics. The dominant topic, with 42 publications, results as Applied physics; from this, the *DR* of his portfolio is 82% (194/236). Figure 3 illustrates the distribution of the indicator for all 947 diversified professors of FIS/01-Experimental physics. The mean value of *DR* is 58.5% (median 60%), the coefficient of variation is 0.269 and min-max interval is 7%-92%.

*Figure 3: Distribution of diversification ratio of professors in FIS/01 (Experimental physics)*

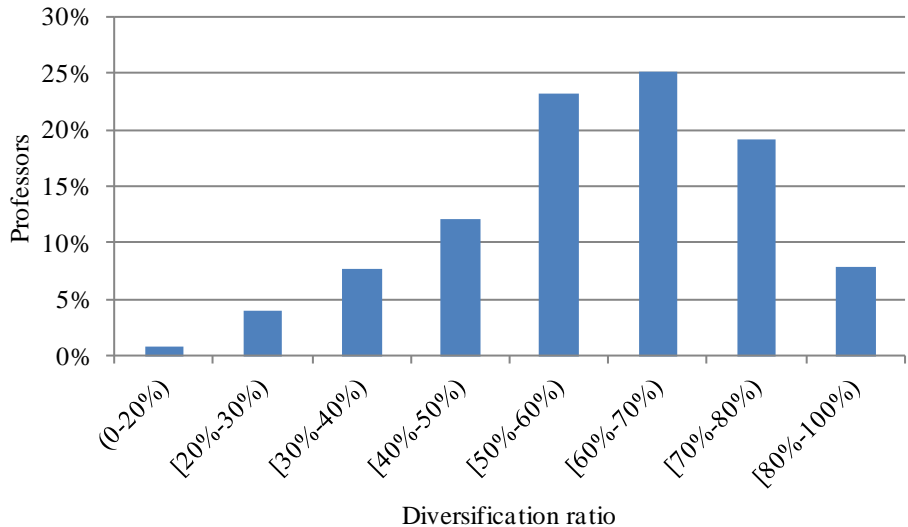

Table 5 provides descriptive statistics of *DR* for professors in all SDSs of Physics. FIS/07 (Applied physics) with 312 professors, shows the highest average *DR,* at 65%. The SDS name itself, "Applied physics (cultural heritage, environment, biology and medicine)", implies ample scope for the field, one where it would be difficult for a specific research topic to "dominate". Still, all the others SDSs show mean *DR* levels above 50%, with the exception of FIS/05 (Astronomy and astrophysics), where 130 professors have an average *DR* of 28%. FIS/05 also stands out, for *DR* coefficient of variation (0.770): in all other SDSs the value never exceeds 0.31, indicating little dispersion around the mean. The broadest range of variation in *DR* (min-max) is observed for professors of FIS/01-Experimental physics (7%-92%).

*Table 5: Descriptive statistics of diversification ratio (DR) for diversified professors in SDSs of Physics*

| SDS† | No of professors | Average DR | Min – Max DR | Var. coeff. |
|---|---|---|---|---|
| FIS/01 | 947 | 59% | 7%-92% | 0.269 |
| FIS/02 | 313 | 53% | 7%-87% | 0.305 |
| FIS/03 | 435 | 62% | 6%-90% | 0.245 |
| FIS/04 | 158 | 56% | 13%-83% | 0.275 |
| FIS/05 | 130 | 28% | 2%-78% | 0.770 |
| FIS/06 | 53 | 57% | 14%-80% | 0.295 |
| FIS/07 | 312 | 65% | 10%-91% | 0.239 |
| FIS/08 | 16 | 51% | 29%-67% | 0.217 |

† *FIS/01, Experimental physics; FIS/02, Theoretical physics, Mathematical models and methods; FIS/03, Material physics; FIS/04, Nuclear and subnuclear physics; FIS/05, Astronomy and astrophysics; FIS/06, Physics for earth and atmospheric sciences; FIS/07, Applied physics (cultural heritage, environment, biology and medicine); FIS/08, Didactics and history of physics*

The results of the analysis repeated for all SDSs are found in SM2. Figure 4 provides the plot of mean *DR* values for all SDSs considered. We note the presence of only two SDSs with mean *DR* greater than 70% (ING-INF/04 Systems and control engineering and ING-INF/06 Electronic and information bioengineering), and six with *DR* less than 40% (FIS/05 Astronomy and astrophysics, MAT/02 Algebra, MED/16 Rheumatology, MED/30 Eye diseases, VET/08 Clinical veterinary medicine and VET/09 Clinical veterinary surgery).

*Figure 4: Distribution of average diversification ratio (DR) for 184 SDSs with at least 10 diversified professors*

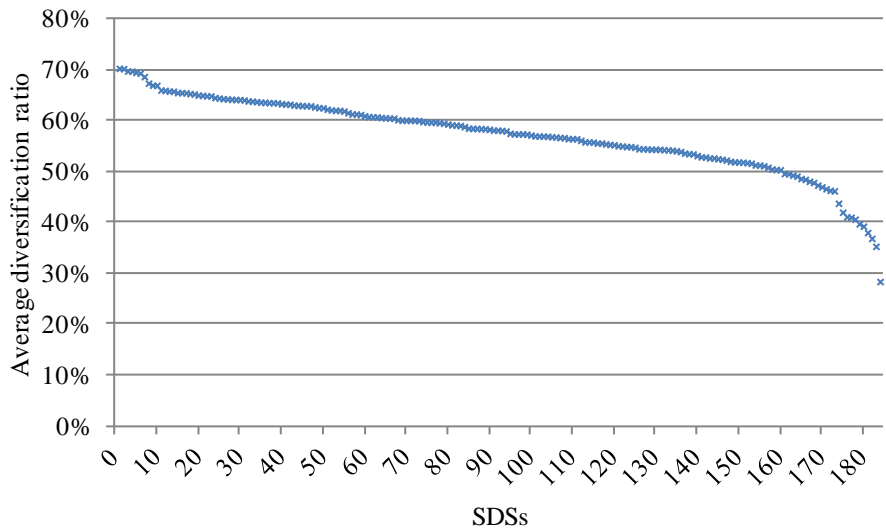

Table 6 presents some statistics concerning distribution of *DR* for professors of all SDSs examined, per UDA. We observe average values very close to each other, over a range from 55% in Earth sciences (UDA 4) to 64% in Industrial and information engineering (UDA 9). In reality the distributions for the individual professors show very pronounced variability, with a range (min-max) between 80 and 90 percentage points within each UDA. We then identify the SDS per UDA with maximum and minimum mean DR: concerning maximums, the values are very similar, between 63% for GEO/05 (Applied geology, of UDA 4) and 70% for ING-INF/04 (Systems and control engineering, of UDA 9). However there is greater variability concerning minimums, with an excursion from 28% in FIS/05 (Astronomy and astrophysics, of UDA 2) to 54% in ICAR/05 (Transport, of UDA 8).

*Table 6: Descriptive statistics of diversification ratio (DR) by UDA*

| UDA† | Average DR | Min-Max DR | Min DR (SDSs average) | Max DR (SDSs average) |
|---|---|---|---|---|
| 1 | 56% | 6%-92% | 35% (MAT/02) | 67% (MAT/09) |
| 2 | 57% | 2%-92% | 28% (FIS/05) | 65% (FIS/07) |
| 3 | 58% | 4%-91% | 51% (CHIM/06) | 66% (CHIM/01) |
| 4 | 55% | 8%-89% | 47% (GEO/10) | 63% (GEO/05) |
| 5 | 62% | 8%-91% | 49% (BIO/04) | 66% (BIO/17) |
| 6 | 57% | 3%-92% | 39% (MED/16) | 70% (MED/01) |
| 7 | 56% | 9%-90% | 37% (VET/08) | 66% (AGR/07) |
| 8 | 60% | 11%-94% | 54% (ICAR/05) | 65% (ICAR/08) |
| 9 | 64% | 3%-92% | 47% (ING-IND/20) | 70% (ING-INF/04) |
| Total | 59% | 2%-94% | 28% (FIS/05) | 70% (ING-INF/04) |

† 1, Mathematics and computer science; 2, Physics; 3, Chemistry; 4, Earth sciences; 5, Biology; 6, Medicine; 7, Agricultural and veterinary sciences; 8, Civil engineering; 9, Industrial and information engineering

### 4.3. Degree of relatedness

In this section we present the results from the analysis concerning *degree of relatedness* of the topics in which the researchers diversify their scientific production. The diversification of production could in fact be confined to a single disciplinary area (mathematics, physics, chemistry, etc.), or spread beyond. The *RR* indicator measures the share of publications that cover topics in the same discipline, out of total. Table 7 returns to our example of the Physics UDA, showing descriptive statistics of *degree of relatedness* for diversified professors of the 8 SDSs.

The percentage of professors who publish in different disciplines varies from a minimum of 50% in FIS/02 (Theoretical physics, mathematical models and methods) to maximum of 100% in FIS/08 (Didactics and history of physics) followed by FIS/07 (Applied physics) with 98%. The results are plausible, considering the cognitive character of the fields: FIS/02 is certainly a more closed and self-contained field than FIS/07, where research is by definition strongly oriented towards applications in different disciplines. Column three in Table 7 shows mean value of *degree of relatedness* for professors of the eight fields: the highest value (79%) is observed in FIS/02 (Theoretical physics, mathematical models and methods); in contrast, FIS/07 (Applied physics), with only 49% of production in the dominant discipline, is again revealed as the most heterogeneous field, in terms of spectrum of scientific activity by scientists practicing. Examining the different SDSs, we observe quite similar distributions of *degree of relatedness,* at least in terms of range of min-max variation. FIS/05 (Astronomy and astrophysics) stands out for quite high minimum value (40%). Concerning maximums, in all the SDSs we observe cases of professors with *degree of relatedness* near 100%, nevertheless only in FIS/01 there is a case of a professor with production limited entirely to a single discipline.

*Table 7: Descriptive statistics of research field relatedness (RR) in the SDSs of Physics*

| SDS† | % of diversified professors publishing in different disciplines | Average RR | Min - Max RR |
|---|---|---|---|
| FIS/01 | 94% | 64% | 18%-100% |
| FIS/02 | 50% | 79% | 25%-98% |
| FIS/03 | 90% | 64% | 19%-97% |
| FIS/04 | 73% | 71% | 25%-99% |
| FIS/05 | 65% | 77% | 40%-98% |
| FIS/06 | 94% | 58% | 25%-95% |
| FIS/07 | 98% | 49% | 17%-98% |
| FIS/08 | 100% | 61% | 33%-96% |

† *FIS/01, Experimental physics; FIS/02, Theoretical physics, Mathematical models and methods; FIS/03, Material physics; FIS/04, Nuclear and subnuclear physics; FIS/05, Astronomy and astrophysics; FIS/06, Physics for earth and atmospheric sciences; FIS/07, Applied physics (cultural heritage, environment, biology and medicine); FIS/08, Didactics and history of physics*

Table 8 offers a summary analysis from all the SDSs under observation; the details for each SDS[9] are available in SM3. The percentage of professors publishing in other disciplines varies from 57% in UDA 4 (Earth sciences) to 93% in UDA 3 (Chemistry). For each UDA we can see the breadth of min-max range of *RR*: the lowest is in Chemistry (15 percentage-points between CHIM/10 Food chemistry and CHIM/11 Chemistry and biotechnology of fermentations) and maximum is in Earth sciences (63 percentage points between SDSs GEO/02 Stratigraphic and sedimentological geology and GEO/06 Mineralogy). Concerning *RR*, the mean values per UDA are quite similar, varying from the 54% of UDAs 3, 5 and 8 (Chemistry, Biology and Civil engineering), to 65% of UDA 7 (Agricultural and veterinary sciences). The last two columns show the SDSs with the minimum and maximum average *RR* for each UDA. Here, BIO/15 (Pharmaceutic biology, in UDA 5) shows the absolute lowest mean *RR* (43%); absolute maximum (79%) is observed in FIS/02 Theoretical physics, mathematical models and methods (UDA 2) and AGR/20 Animal husbandry (UDA 7). The former case (BIO/15) reveals a field where research by nature evidently overlaps with different disciplines; the second (FIS/02 and AGR/20), confirms the particular character of fields with clearly marked disciplinary boundaries.

*Table 8: Descriptive statistics of the degree of relatedness (RR) UDA*

| | % of diversified professors publishing in different disciplines | | Relatedness ratio (%) | |
|---|---|---|---|---|
| UDA† | Average | Min-Max | Average | Min-Max |
| 1 | 75 | 36 (MAT/02)-98 (MAT/09) | 63 | 49 (MAT/09)-69 (MAT/02) |
| 2 | 85 | 50 (FIS/02)-100 (FIS/08) | 64 | 49 (FIS/07)-79 (FIS/02) |
| 3 | 93 | 84 (CHIM/11)-99 (CHIM/10) | 54 | 44 (CHIM/12)-65 (CHIM/11) |
| 4 | 57 | 30 (GEO/02)-93 (GEO/06) | 64 | 53 (GEO/09)-76 (GEO/03) |
| 5 | 91 | 57 (BIO/04)-99 (BIO/15) | 54 | 43 (BIO/15)-74 (BIO/04) |
| 6 | 87 | 58 (MED/20)-100 (MED/45) | 62 | 50 (MED/42)-77 (MED/23) |
| 7 | 74 | 40 (AGR/20)-98 (VET/06) | 65 | 51 (AGR/01)-79 (AGR/20) |
| 8 | 90 | 79 (ICAR/09)-95 (ICAR/08) | 54 | 47 (ICAR/08)-64 (ICAR/09) |
| 9 | 85 | 70 (ING-INF/05)-100 (ING-IND/12) | 62 | 44 (ING-IND/23)-78 (ING-INF/07) |
| Total | 85 | 30 (GEO/02)-100 (various) | 60 | 43 (BIO/15)-79 (FIS/02) |

† *1, Mathematics and computer science; 2, Physics; 3, Chemistry; 4, Earth sciences; 5, Biology; 6, Medicine; 7, Agricultural and veterinary sciences; 8, Civil engineering; 9, Industrial and information engineering*

---

[9] Here too, we omit the SDSs (8 in all) with less than 10 diversified professors.

## 5. Conclusions

Diversification in research activity is often motivated by the opportunity to apply one's own competencies in areas different from that of specialization. Curiosity, typical of all explorers, tempts researchers towards little-investigated fields. At the same time, the increasing complexity of phenomena under investigation often requires the participation of different specialists. This is reflected in the increasing number of coauthored publications and coauthors per publication. In this work we have investigated the research-diversification behavior of academics, through classification by subject category of their publication over the period 2004-2008. We defined "diversified researchers" as those whose publications fall in more than one subject category or combination, and then measured the share of diversified authors out of total in every field and discipline. In almost all fields, the vast majority of academics diversify their scientific production. We analyzed the nature of such diversification along three dimensions: extent of diversification, intensity of diversification, and degree of relatedness of topics in which the academic diversifies. Each dimension was measured using a specific indicator. The results obtained from such analysis obviously depend on both the breadth of time for observation of scientific production and the scheme applied for sectoral classification of the scientific production. Given this, the importance of the measures is not in their absolute value, rather in the relative values between fields (SDSs) to which the researchers belong.

We found that the extent of diversification (number of topics covered by the scientist's research portfolio, different than the dominant one) varies among fields within individual disciplines and among disciplines, and is highly correlated to the intensity of publication. It is lowest in Mathematics and highest in Chemistry. Variations among fields and disciplines are also observed using the other indicators. Intensity of diversification (share of publications outside the academic's dominant topic of research) is lowest in Earth sciences and highest in Industrial and information engineering, but differences are generally less notable among disciplines. Degree of relatedness (share of publications falling in the same discipline) is lowest in Earth sciences and highest in Chemistry.

The next step will be to discover if this diversification pays off, meaning if publications outside the author's dominant field of specialization have a higher relative impact than the others. Our next study on the theme will be dedicated to this question.

**Supplementary material**

SM 1 - Extent of research diversification of professors by SDS
SM 2 - Intensity of research diversification of professors by SDS
SM 3 - Degree of relatedness of professors by SDS